\documentstyle[aps,prbbib,twocolumn,epsf]{revtex}

\begin{document}
\draft
\title{Parametric pumping at finite frequency}

\author{Baigeng Wang$^1$, Jian Wang $^1$ and Hong Guo$^{2}$}
\address{1. Department of Physics, The University of Hong Kong, 
Pokfulam Road, Hong Kong, China\\
2. Center for the Physics of Materials and Department 
of Physics, McGill University, Montreal, PQ, Canada H3A 2T8\\}
\maketitle

\begin{abstract}
We report on a first principles theory for analyzing the parametric 
electron pump at a finite frequency. The pump is controlled
by two pumping parameters with phase difference $\phi$. In the 
zero frequency limit, our theory predicts the well known
result that the pumped current is proportional to $\sin\phi$. 
For the more general situation of a finite frequency, 
our theory predicts a non-vanishing pumped current even when the
two driving forces are in phase, in agreement with the recent 
experimental results. We present the physical mechanism behind the 
nonzero pumped current at $\phi=0$, which we found to be
due to photon-assisted processes.
\end{abstract}


Parametric electron 
pump\cite{altshuler,switkes,brouwer,zhou,shutenko,aleiner,vavilov}
is an interesting device which delivers a 
finite DC current to the outside world at {\it zero} bias
potential, by cyclic variations of two device control parameters.
Recently, an adiabatic quantum electron pump was reported in an open 
quantum dot where the pumping signal was produced in response to the cyclic 
deformation of the confining potential\cite{switkes}. It was found that 
the pumping signal, $V_{dot}(\phi)$, is sinusoidal in the phase difference 
$\phi$ between the two deforming potentials in the weak pumping regime, and 
it becomes non-sinusoidal in the strong pumping regime. The amplitude of the 
pumping signal increases linearly with the frequency of the deformation. 
Most notable, however, was the data\cite{switkes} showing 
$V_{dot}(0) \neq 0$ 
significantly at strong pumping, whereas $V_{dot}(\pi)\approx 0$ for 
all pumping strength. Even in the weaker pumping 
regime, small deviations from $V_{dot}(0)=0$ could already be
seen\cite{switkes}. The traditional and successful parametric pumping 
theory\cite{brouwer,zhou}, valid in the adiabatic regime and up to first 
order in frequency, requires two pumping parameters which traverse in 
a closed path in parameter space in each cyclic period and the pumping 
signal is proportional to the enclosed area by the path. Accordingly,
if the two pumping parameters are in phase so that the enclosed 
area is zero, the pumping signal $V_{dot}(0)$ should vanish. It is extremely
puzzling that the experimental data\cite{switkes} consistently
showed $V_{dot}(0) \neq 0$, 
a fact which has not been understood so far,
and, clearly, it calls for the development of a first principle theory going
beyond the adiabatic regime and low frequencies.

It is the purpose of this letter to present a theory for parametric pumping
which is valid at {\it finite} frequency. Using this theory
we investigate the frequency dependence of the pumping current   
$I_p(\phi) = G_{dot} V_{dot}(\phi )$ with $G_{dot}$ a constant, and it
allows us to understand why it is possible to have a pumping signal even   
when the pumping forces are exactly in-phase.
When frequency is low, our theory recovers the traditional 
results\cite{brouwer,zhou}. As the frequency increases, 
we predict a nonzero $I_p(0)$ which is a consequence of photon 
assisted processes, and our theory also predicts $I_p(\pi) \approx 0$.
These results allows us to reach the conclusion
that the experimental data\cite{switkes} showing $V_{dot}(0) \neq 0$ 
and $V_{dot}(\pi)\approx 0$, are generic nonlinear transport features 
of parametric pumps at finite frequency. Furthermore, our theory 
predicts that at very large frequencies $I_p(\pi)$ should start 
to deviate from zero; and most intriguing is the natural theoretical
outcome that only one periodic deforming potential can produce a 
pumping signal at finite frequency. The latter is due to the fact
that a finite frequency provides extra degrees of freedom through
photon assisted processes that is capable of playing the role of 
a second pumping parameter. These new predictions should be testable
experimentally.

We start by considering a parametric pump which consists of a 
coherence quantum scattering region attached to two ideal 
leads $L,R$. The leads maintain {\it identical} electrochemical 
potential, {\it i.e.}, $\mu _{L}=\mu _{R}=\mu $. The Hamiltonian of 
this system is\cite{foot10}
\begin{eqnarray}
%H_{0}&=&\sum_{k,\alpha =L,R}\epsilon _{k}C_{k\alpha }^{+}C_{k\alpha
%}+H_{cen}\{d_{i},d_{i}^{+}\} \nonumber \\
%&+&\sum_{k,\alpha ,i}(T_{k,\alpha ,i}C_{k\alpha}^{+}d_{i}+c.c.)
H_{0}&=&\sum_{k,\alpha =L,R}\epsilon _{k}C_{k\alpha }^{+}C_{k\alpha
}+\sum_i [\epsilon_i(t) d^+_i d_i \nonumber \\
&+&(t_i d^+_i d_{i+1} +c.c.)]
+\sum_{k,\alpha}(T_{k,\alpha}C_{k\alpha}^{+}d_{j}+c.c.)
\end{eqnarray}
where $C_{k\alpha }$ and $d_{i}$ are annihilation operators of electrons 
for the $\alpha$ lead and the scattering region at site $i$, 
respectively\cite{foot6}.
The three terms describe the leads, the scattering region, and the 
the coupling between the leads and the scattering region with the hopping 
matrix $T_{k,\alpha}$. In the last term, $j=1/N$ for $\alpha=L/R$.
The parametric pump works by cyclic deformations of potential at two 
different pumping sites $i$ and $j$ in the scattering region,
$V_{i/j}(t)=V_{i/j}\cos (\omega t+\varphi_{i/j})$
with $\varphi_{i/j}$ the phase of the pumping force.

Neglecting interaction between electrons in the ideal leads, the Keldysh 
nonequilibrium Green's function (NEGF) theory gives the following standard 
expression for the time dependent current\cite{jauho} ($\hbar=1$),
\begin{eqnarray}
I_{L}(t)&=&-q\int_{-\infty }^{t}dt_{1}[G^{r}_{11}(t,t_{1})\Sigma _L
^{<}(t_{1},t) \nonumber \\
&+&G^{<}_{11}(t,t_{1})\Sigma_L^{a}(t_{1},t)+c.c.]
\label{current1}
\end{eqnarray}
where the Green's functions $G^{r,a,<}$ and the self-energy 
$\Sigma_{\alpha}$ are defined as in the usual manner\cite{datta}.

It is tedious but straightforward to evaluate the Green's functions
by iterating the equation of motion\cite{jauho,datta}, and for our purpose
it is adequate to calculate them to second order in the pumping potential 
$V_{i/j}(t)$ which gives the average pumped current to the second order.
We note that the next higher non-vanishing order is the fourth.
After all the Green's functions are obtained, the average pumped current is
calculated from Eq.(\ref{current1}) by integrating time over one pumping 
cycle. We obtain,
\begin{equation}
I_p\equiv <I_L> = I_{ii}+I_{jj}+I_{ij}+I_{ji}
\label{eq12}
\end{equation}
where 
\begin{eqnarray}
&&I_{ij} = -\frac{iqV_{i}V_{j}}{8}\int \frac{dE}{2\pi}\Gamma_{L}
G_{1i}^{0r}G_{j1}^{0a} \{(f-f_{-}) \nonumber \\
&\times&
e^{i\phi}[G_{ij}^{0r-}-G_{ij}^{0a-}] 
+(f-f_{+})e^{-i\phi}[G_{ij}^{0r+}-G_{ij}^{0a+}] \}
\label{final}
\end{eqnarray}
where $f\equiv f(E)$ and $f_{\pm}\equiv f(E\pm\omega )$ are the Fermi 
distribution functions; $\Gamma _{L} =-2Im[\Sigma_{L}^{r}]$ is the line 
width function; $\phi\equiv \varphi_{j}-\varphi_{i}$ is the phase 
difference between the two pumping forces. Finally, $G^{0r} \equiv G^{0r}(E)$ 
and $G^{0r\pm} \equiv G^{0r}(E\pm\omega)$ are the retarded Green's functions
when there is no pumping force. In Eq.(\ref{final}), $I_{ii}$ is obtained 
by setting $j=i$. Eqs.(\ref{eq12},\ref{final}) are the main result of this 
work.

Before we discuss Eqs.(\ref{eq12},\ref{final}) in connection to the
experimental data of Ref.\onlinecite{switkes}, let's first examine the low 
frequency limit of these results and show that the conventional parametric
pumping theory is recovered\cite{brouwer,zhou}. We expand Eq.(\ref{final}) 
in powers of $\omega$ and only keep the {\it linear} term, this gives the 
adiabatic current\cite{brouwer,zhou}. Note that $I_{ii}$ and $I_{jj}$ are of 
the second order in frequency, Eq.(\ref{final}) reduces to
\begin{eqnarray}
I_p &=& \frac{iqV_{i}V_{j}\omega}{4}\int \frac{dE}{2\pi} \partial_E f 
\sin(\phi) \Gamma_{L} \nonumber \\
&\times& [\frac{\partial G_{11}^{0r}}{\partial V_i} 
\Gamma_L \frac{\partial G_{11}^{0a}}{\partial V_j}  
+ \frac{\partial G_{1N}^{0r}}{\partial V_i} 
\Gamma_R \frac{\partial G_{N1}^{0a}}{\partial V_j}] + c.c.
\label{scatter}
\end{eqnarray}
where we have used the relation $G^r-G^a = -i G^r \Gamma G^a$ 
and\cite{gasparian} $G^r_{\alpha i} G^r_{i \beta} = \partial 
G^r_{\alpha \beta}/\partial V_i$. Using the Fisher-Lee 
relation\cite{fisher,datta} $S = -I + i \Gamma^{1/2} G^r \Gamma^{1/2}$, 
Eq.(\ref{scatter}) reduces to exactly the same expression as that obtained 
from scattering matrix theory\cite{brouwer}. Due to the factor $\sin(\phi)$ 
in the low frequency result (\ref{scatter}), one obtains the familiar 
outcome\cite{brouwer} that $I_p(\phi)=0$ at both $\phi =0$ and $\phi = \pi$.

Clearly, the low frequency result (\ref{scatter}) does not explain the 
experimental result\cite{switkes} of $I_p(\phi=0) \neq 0$, we need to
investigate the full result Eq.(\ref{final}) at a finite frequency.
The first term on the right hand side of (\ref{final}) has a clear
physical meaning: it represents the photon absorption process indicated by
the factor $\exp(i\omega t +i\phi)$. Similarly, the second term corresponds 
to the photon emission process with factor $\exp(-i\omega t -i\phi)$.
These two competing processes tend to cancel each other in the expression 
of pumped current. These photon assisted processes are essential to 
understand the experimental 
finding\cite{switkes} that the pumped signal is nonzero at $\phi=0$.
To see this clearly, we rewrite Eq.(\ref{eq12}) into the following form
\begin{eqnarray}
I_p &=& -\frac{q}{8} \int \frac{dE}{2\pi} \left\{ (f-f_{-})
\right.\nonumber \\
&\times& \left[ |g^1_i(E,E-\omega) + e^{-i\phi} g^1_j(E,E-\omega)|^2 
\right.\nonumber \\
&+& \left.|g^N_i(E,E-\omega) + e^{-i\phi} g^N_j(E,E-\omega)|^2 \right]
\nonumber \\
&+& (f-f_{+}) \left[|g^1_i(E,E+\omega) + e^{i\phi} g^1_j(E,E+\omega)|^2 
\right.\nonumber \\
&+& \left. \left.|g^N_i(E,E+\omega) + e^{i\phi} g^N_j(E,E+\omega)|^2 \right]
\right\}
\label{final1}
\end{eqnarray}
where $g^\alpha_i(E,E-\omega) \equiv \sqrt{\Gamma_1 \Gamma_\alpha}
[G_{1i}^{0r}(E)] V_i [G_{i\alpha}^{0r}(E-\omega)]$ with $\alpha = 1, N$ 
indicating the left and right positions where scattering region is contacted
by the leads, so that $\Gamma_1\equiv \Gamma_L$ and $\Gamma_N\equiv \Gamma_R$. 
The propagator $G_{mn}^{0r}$ describes the free motion of an electron from 
position $n$ to position $m$ in the device. Therefore, the quantity 
$g^1_i(E,E-\omega)$ describes the following process: a charge carrier with 
energy $E-\omega$ enters the device from the left lead, it absorbs a photon 
with frequency $\omega$ at site $i$, and then exits from the left lead with 
energy $E$. This process is represented by the Feynman diagram in the inset 
of Fig.(1). Similarly, $g^1_j(E,E-\omega)$ describes exactly the same
process except that the electron absorbs a photon at position $j$.
Now the physics is transparent: the first term in Eq.(\ref{final1}) 
represents an interference of the photon-absorption processes happening
at positions $i,j$, {\it i.e.} the interference of the two processes
in the Feynman diagram of Fig.(1). Therefore, when $\phi=0$, we have a
constructive interference so that the $g^1_i(E,E-\omega)$ and the 
$g^1_j(E,E-\omega)$ process add up (the factor $\exp(-i\phi)=+1$). When
$\phi=\pi$, there is a destructive interference in which the two
photon-absorption processes cancel to a large extent (the factor
$\exp(-i\phi)=-1$). Exactly the same can be said for the three other terms 
of Eq.(\ref{final1}): the term involving $g^N_i(E,E-\omega)$ describes 
photon-absorption process with electrons entering the device at the right 
lead and exits from the left; and the terms involving $g^1_i(E,E+\omega)$ 
and $g^N_i(E,E+\omega)$ describe photon-emission processes. 

In addition to the interference effects, the pumped current is also affected
by a competition between photon emission and absorption, marked by the 
$(f-f_{-})$ term for absorption and the $(f-f_{+})$ term for emission, 
in Eq.(\ref{final}). 
The combined effect, competition plus interference,
can be clearly seen by expanding Eq.(\ref{final1}) to order $O(\omega^2)$ 
which produces a complex expression for general $\phi$, but if we 
concentrate only on $\phi=0$ or $\pi$, the result is much simpler and 
physically transparent,
\begin{eqnarray}
I_p &=& \frac{q}{16\pi} \omega^2 (\partial_{E_1} -\partial_{E_2})
\left [ |g^1_i(E_1,E_2) +e^{i\phi} g^1_j(E_1,E_2)|^2  \right.\nonumber \\
&+& \left. |g^N_i(E_1,E_2) + e^{i\phi} 
g^N_j(E_1,E_2)|^2\right] 
\end{eqnarray}
where one sets $E_1=E_2=E$ after taking the derivative and the
temperature is set to zero. Hence, at
$\phi=0$ the constructive interference gives larger current than
at $\phi=\pi$ (see also the right inset of Fig.2 described below). 
The behavior of the pumped current for general $\phi$ 
is the result of the interplay between the photon assisted processes 
and interference processes.

The above physical picture allows us to unambiguously conclude that 
the behavior of pumped current depends on a combination of
interference and competition of photon assisted processes. 
The competition between absorption and emission sets an overall
magnitude for the pumped current at each $\phi$.
Destructive interference occurs at $\phi=\pi$, therefore
$I_p(\pi)\approx 0$; constructive interference occurs at $\phi=0$, 
giving rise to a non-vanishing $I_p(0)$. To demonstrate this, 
in the right inset of Fig.2, we plot the magnitude of pumped current versus 
$\phi$ at a finite frequency due to the contributions of the photon 
absorption (solid line, first two terms of Eq.(\ref{final1})) and photon 
emission (dotted line, the last two terms of Eq.(\ref{final1})) processes, 
respectively. Taking into account of absorption and emission processes, 
$I_p(0)$ is significantly larger than $I_p(\pi)$\cite{foot3}. The result 
in this inset is obtained from an one-dimensional model which we now 
discuss in more detail.

Our discussion so far is completely general on the general result 
Eq.(\ref{final}). In the following we investigate a model in a more 
specific manner by applying Eq.(\ref{final}) to a double barrier quantum 
structure which we model by potential $U(x)=V_0 
\delta (x)+V_0 \delta (x-a)$ where $V_0$ is the barrier height and 
$a$ the barrier separation\cite{foot}. For this system the Green's function 
$G(x,x')$ can be calculated exactly\cite{yip}. We choose the pumping force 
as $V(t)= V_i\delta(x-x_i)\sin(\omega t)+ V_j\delta(x-x_j)
\sin(\omega t+\phi)$. With this specific pump model, Eq.(\ref{final}) can 
be evaluated numerically without difficulty\cite{foot4}.

In Fig.1, we plot the pumped current $I_p$ at zero temperature versus the 
Fermi energy at different frequencies with phase difference $\phi=\pi/2$. 
For comparison, we also plot transmission coefficient (long dashed
line). The peak in 
transmission coefficient indicates a quantum resonance mediated by the 
resonance state in the double barrier. Clearly, the pumped current $I_p$ 
also shows a resonance behavior, peaked at the same resonance state and
is largely suppressed away from it. As the frequency is increased, the 
pumped current reverses the sign and the peak is shifted slightly.
The general feature of $I_p$ versus phase difference $\phi$ can be
obtained from Eq.(\ref{final}) which can be rewritten as $I_p = c_1 + c_2 
\sin \phi + c_3 \cos \phi = c_1 + c_4 \sin (\phi + \phi_0)$, where $c_i$s 
and $\phi_0$ are constants. This indicates that $I_p$ has a sinusoidal 
behavior in $\phi$. The sinusoidal behavior is a direct consequence of the
fact that our theory is to the second order of the pumping 
amplitude\cite{foot1}.

Fig.2 presents $I_p$ as a function of $\phi$ for several different 
frequencies at resonance where the transmission coefficient is unity. 
We notice that at very low frequency (solid line) $I_p \sim \sin(\phi)$.
As the frequency increases, the amplitude of $I_p$ also increases. At the 
same time, the entire $I_p$-$\phi$ curve is shifted upward although leaving
$I_p(\phi=\pi)$ largely unchanged. However, $I_p(\phi=0)$ is seen to 
increase significantly for the curve with $\omega=0.13$. 
As already discussed above, these features are
due to interference of the photon assisted processes, and they are in
very good agreement with the experimental observation\cite{switkes}. 
The left inset of Fig.2 shows $I_p$ as a function of frequency for several 
phase differences.
We observe that at small frequencies $I_p\approx 0$ at $\phi=0$ and $\pi$. 
For $\phi=\pi/2$ and $3/2\pi$, $I_p$ have similar values. At larger 
frequencies, $I_p$ increases in different fashion. For instance, $I_p$
at $\phi=0$ increases slowly at small frequency and then has a linear
behavior for larger frequencies. At $\phi=\pi/2$, $I_p$ increases initially,
reaches a maximum, and then decreases slowly. At $\phi=\pi$, $I_p$ is very 
small but nonzero and increases linearly with a much smaller slope. 
Whereas $I_p$ at $\phi=3\pi/2$ is the largest among all the curves. 
This suggests that one should operate the parametric electron pump 
at $\phi=3\pi/2$ to achieve the maximum pumped current.

So far, we have explained the physics behind the peculiar experimental
finding\cite{switkes} of $I_p(\phi=0) \neq 0$: it is due to photon assisted
processes which is a nonlinear phenomenon. However, when the two pumping 
forces act exactly in-phase, nothing distinguishes them (except that 
they act at different positions of the pump). Therefore, it is extremely
interesting to ask: can one operate a pump with only one pumping parameter ?
To check if this is possible, we set $V_j=0$ in Eq.(\ref{eq12}) and 
notice that the resulting $I_p$ still remains finite due to the first term
of Eq.(\ref{eq12}). Fig.3 plots this $I_p$ versus 
energy at different frequencies. Again, $I_p$ is peaked near the resonant 
point. As the frequency increases, the amplitude of $I_p$ also 
increases and its peak shifts towards larger energy due to photon assisted
process. The inset of Fig.3 shows $I_p$ vs the position of the single
pumping site at different frequencies for an energy the resonance. 
Due to the symmetry of the system, $I_p$ is antisymmetric across the
center of the double barrier. It is surprising that the magnitude of 
the pumped current has the same order of magnitude as that pumped by 
two driving forces (see long dashed line in the inset of Fig.2).  
The reason that a pump can operate with only one external driving force is 
directly related to photon assisted processes which happen at a finite 
frequency. It will be interesting to test this prediction by further 
experimental work.

In summary, we have developed a parametric pumping theory valid at finite 
frequency and it shows that the pumped signal for two pumping parameters 
in-phase can deviate from zero due to photon assisted process at finite 
frequency. This explains the anomaly at $\phi=0$ observed experimentally. 
This theory also suggests that even with one pumping parameter, it is 
possible to produce a pumped signal at finite frequency whose amplitude 
is in the same order of magnitude of that of two pumping parameters.

After the paper was submitted, we were made aware of a recent work by 
Brouwer\cite{brouwer1} which examined magnetic field symmetry of
the pump. It suggested that rectification effect might be important
in understanding the $\phi=0$ anomaly.

\noindent
{\bf Acknowledgements.}
We gratefully acknowledge support by a RGC grant from the SAR Government of 
Hong Kong under grant number HKU 7215/99P. H.G. is supported by NSERC of 
Canada and FCAR of Qu\'ebec.

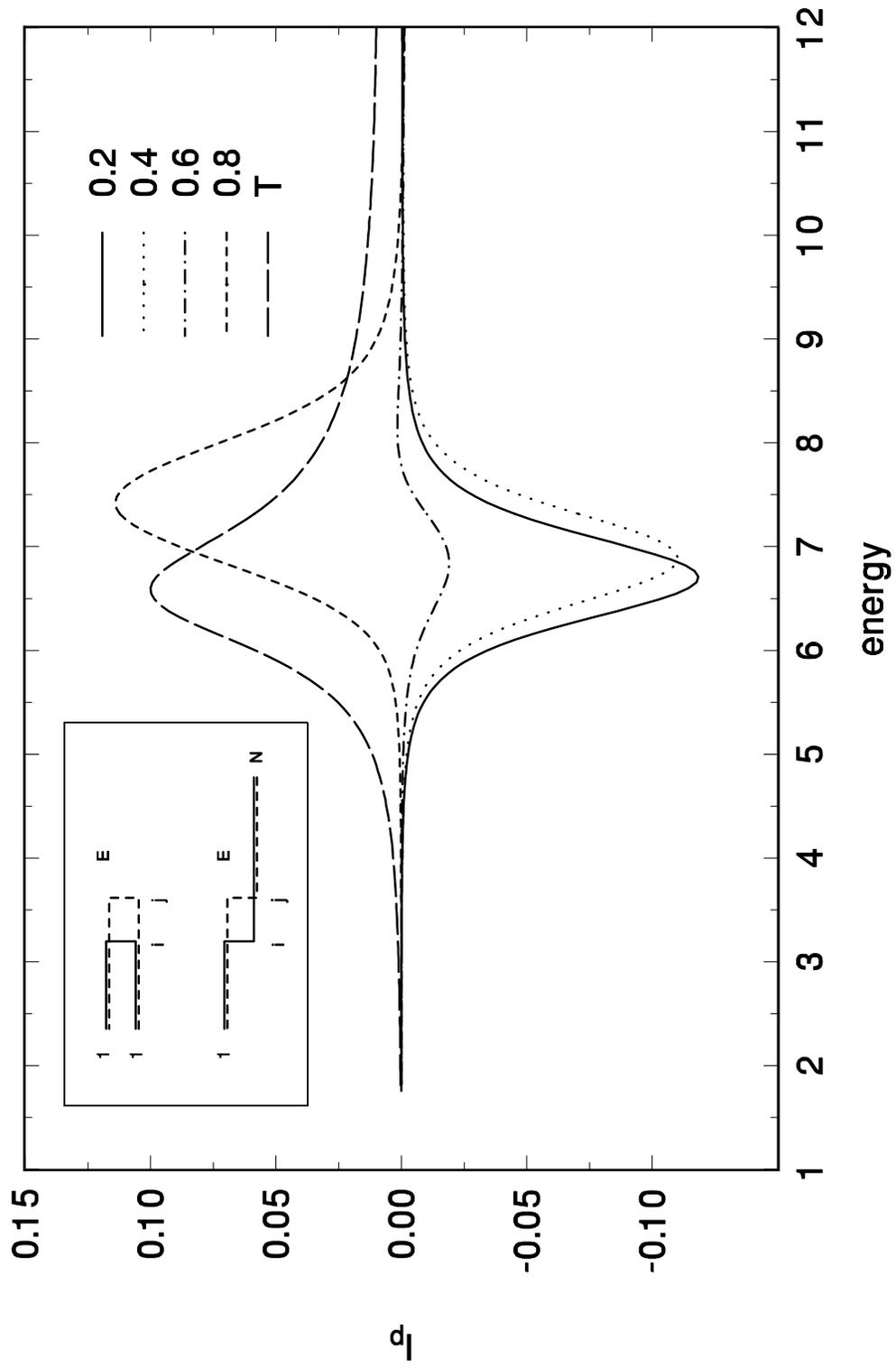
\begin{figure}
\caption{
Pumped current (solid line) and transmission coefficient $T$ (long dashed 
line) versus energy with $\phi=\pi/2$ at $\omega=0.2, 0.4, 0.6, 0.8$. 
For illustrating purpose, we shifted $T$ by multipling a factor 0.1. 
Inset: Feynman diagrams corresponds to Eq.(7).
In Fig.1 and Fig.2, we have set $x_i = 0.02 a$, $x_j = 0.25 a$,
$V_0=79.2$, $V_i=V_j=1$. 
}
\end{figure}

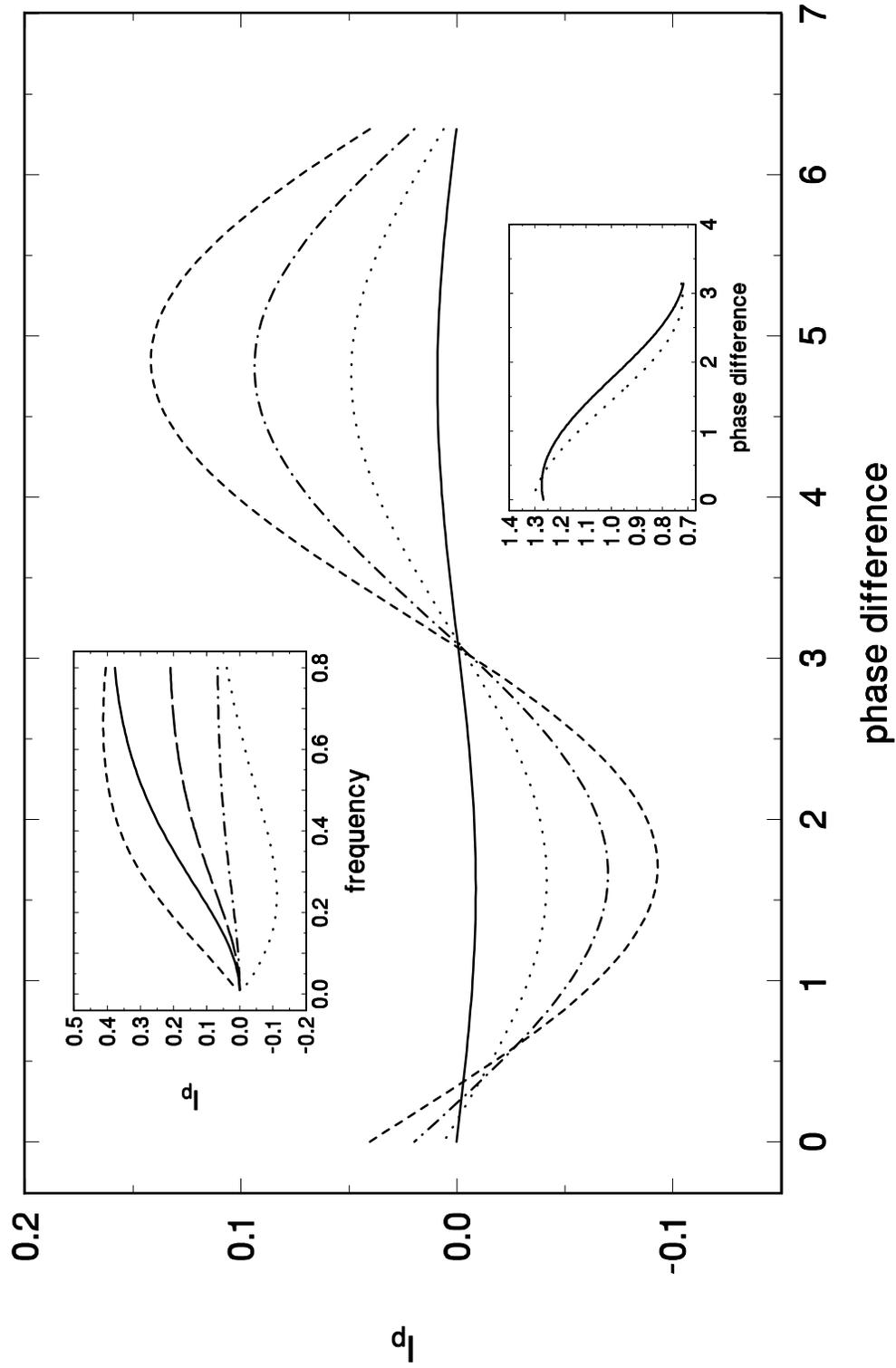
\begin{figure}
\caption{
$I_p$ versus $\phi$ for different $\omega$ at resonant point. Solid line: 
$\omega=0.01$; dotted line: $\omega=0.05$; dash-dotted line: $\omega=0.09$; 
dashed line: $\omega=0.13$. Left inset: $I_p$ versus $\omega$ for 
different values of $\phi$ at resonant point. Solid line: $\phi=0$; 
dotted line: $\phi=\pi/2$; dash-dotted line: $\phi=\pi$; dashed line: 
$\phi=3\pi/2$. The long dashed line is $I_p$ vs $\omega$ for single pumping 
parameter at position $x=0.2a$. Right inset: the magnitude of pumped current 
at resonant point due to contributions of the first two terms (solid line) 
and due to the last two terms (dotted line) in Eq.(7), for $\omega=0.13$.
}
\end{figure}

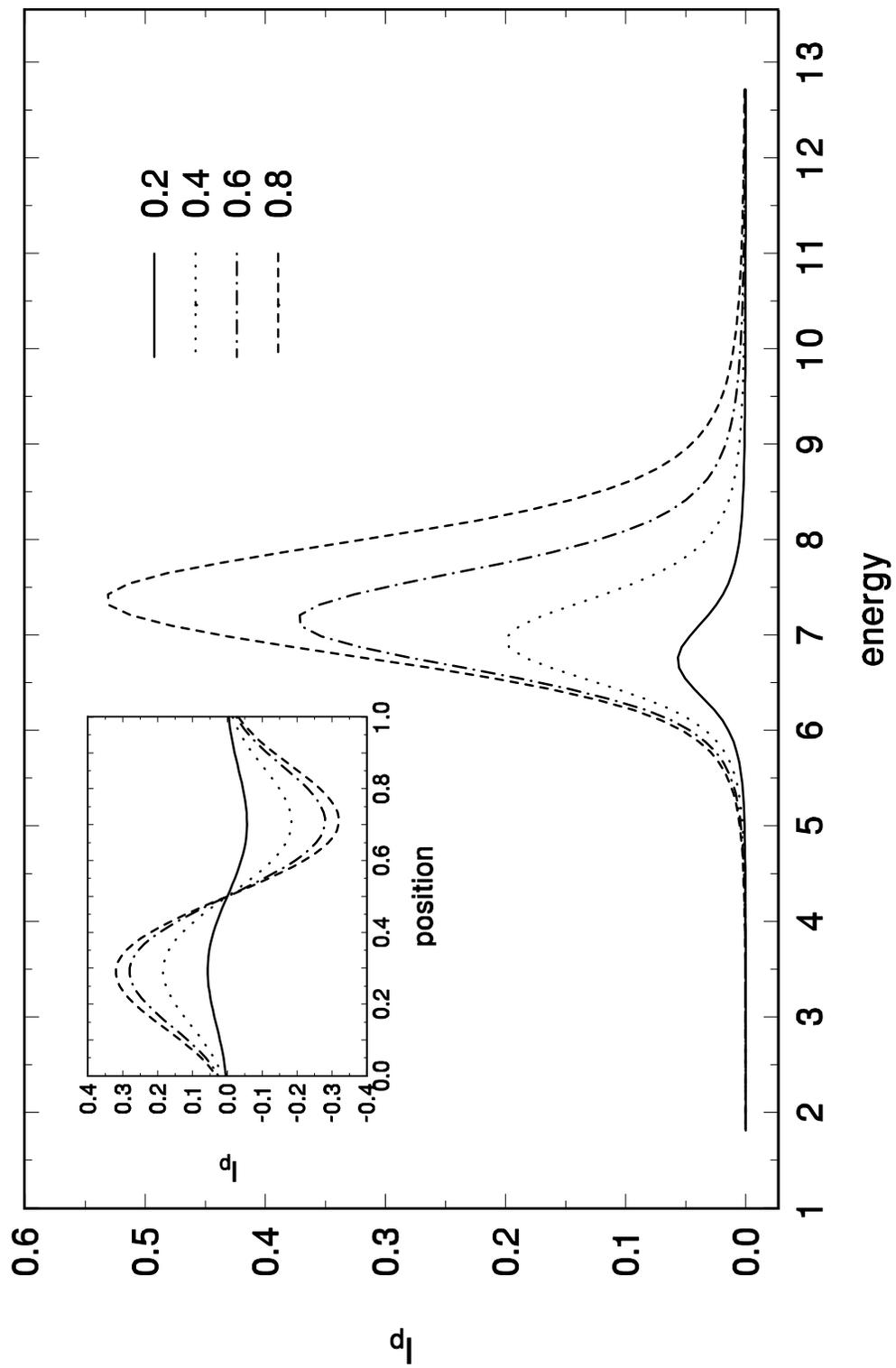
\begin{figure}
\caption{
$I_p$ versus energy at $\omega=0.2,0.4,0.6,0.8$ near resonant point with 
a single pumping parameter. The position of the pumping parameter is 
at $x=0.2a$. Inset: $I_p$ versus position (in unit of $a$) of the pumping 
parameter at $\omega=0.2,0.4,0.6,0.8$. 
}
\end{figure}


\begin{thebibliography}{00}
\bibitem{foot10}
We focus on developing a nonadiabatic theory which is crucial in
understanding the $\phi=0$ experimental anomaly, rather on the chaotic 
nature of the quantum dot in the pump of Ref.\onlinecite{switkes}.
\bibitem{foot6}
Since the pumping potential in the scattering region is
position dependent, it is easiest to work in a real space tight-binding 
representation rather than in momentum space. The scattering region
has $N$ sites. In the numerics we choose $N=400$ to make sure the 
convergence of results.
\bibitem{foot3}
If $x_i$ is closer to $x_j$, the interference effect will be drastically
enhanced. For instance, we have 
$I_p(0)/I_p(\pi) \sim 130$ for $x_i=0.13 a$ and $x_j=0.25 a$ as
compared with $I_p(0)/I_p(\pi) \sim 5$ for $x_i=0.02 a$ and $x_j=0.25 a$ 
shown in the right inset of Fig.2.
\bibitem{foot}
If the double barrier is asymmetric, the pumped current will be 
suppressed.
\bibitem{foot4}
In the numerics we set units by $\hbar=2m=1$.
For a system fabricated by GaAs with $a=1.0\mu m$ 
which corresponds to the linear dimension of the quantum dot in the 
experiment of Ref.\onlinecite{switkes}, our energy unit is $E=2.3\mu eV$.
\end{thebibliography}
\end{document}